\documentclass[%
 reprint,
 amsmath,amssymb,
 aps,
]{revtex4-2}

\usepackage{graphicx}
\usepackage{dcolumn}
\usepackage{bm}

\begin{document}

\preprint{APS/123-QED}

\title{Gauging in Parameter Space: A Top-Down Perspective}

\author{Xingyang Yu$^1$}
\email{xingyangy@vt.edu}
\affiliation{$^1$
 Physics Department, Robeson Hall, Virginia Tech,
 Blacksburg, VA 24061, USA
}

\begin{abstract}
In this paper, we discuss a novel top-down perspective on gauging parameters in quantum field theories (QFTs) by promoting them to partially dynamical fields. Through a generalized notion of symmetry theories, we explore the consequences of this promotion, revealing new topological defects, decompositions as well as generalized symmetry structures across various dimensions. We provide a systematic top-down derivation of symmetry theories for parameters and discuss how special types of branes, such as Euclidean branes and fluxbranes are related to gauging in the parameter space in QFTs. We illustrate our theoretical framework with two key examples: the modified instanton sum in 4D $\mathcal{N}=4$ super Yang-Mills theory and the gauging of gauge ranks in 3D ABJ(M) theories. 
\end{abstract}

\maketitle


\section{\label{sec:intro}Introduction}
In quantum field theories (QFTs), parameters such as coupling constants are typically treated as fixed, background values that define the dynamics of the theory. One useful trick is to promote coupling constants as background fields, in order to investigate the dynamics of the theory.  This approach is standard in many contexts, where coupling constants are treated as fields that vary over spacetime (see, e.g., \cite{Seiberg:1994bp}). In this work, we discuss a new perspective in which these parameters are promoted to partially dynamical fields, fluctuating in a controlled manner but without the introduction of kinetic terms. This approach reveals new topological defects and generalized symmetry structures, revealing deeper topological layers in QFTs.

A significant consequence of this promotion is the possibility of \emph{decomposition} \cite{Hellerman:2006zs, Sharpe:2022ene}, where the original QFT decomposes into a sum over different `universes' \cite{Tanizaki:2019rbk, Komargodski:2020mxz}, each corresponding to a distinct value of the promoted parameter. From the perspective of generalized global symmetries \cite{Gaiotto:2014kfa}, this decomposition corresponds to a $(D-1)$-form symmetry in a QFT defined on a 
$D$-dimensional spacetime. This symmetry is associated with charges carried by domain walls that separate universes. A notable aspect of these domain walls is that their tensions are infinite, in contrast to the finite tensions typically found in domain walls associated with superselection sectors arising from spontaneous symmetry breaking \cite{Komargodski:2020mxz}. The charges associated with these infinitely heavy domain walls are measured by topological local operators, which are constructed from fields that emerge from the promoted parameters.

This promotion can be understood as gauging a `$(-1)$-form symmetry'—a generalization of symmetry concepts to include parameters. In this framework, anomalies and gauging can be extended into the parameter space, as explored in various contexts (see, e.g., \cite{Seiberg:2010qd, Cordova:2019jnf, Cordova:2019uob, Sharpe:2019ddn, Vandermeulen:2022edk, Aloni:2024jpb, Brennan:2024tlw}). Intuitively, the parameter can be viewed as a background gauge field for a $(-1)$-form symmetry. Upon gauging the $(-1)$-form symmetry, a dual $(D-1)$-form symmetry emerges as a quantum symmetry. This emergence requires gauging a finite subgroup—or more precisely, a finite subset—of the parameter space, consistent with our approach of making the parameters partially dynamical, without introducing kinetic terms.

While aspects of these phenomena have been discussed in various contexts, the existing treatments have been purely field-theoretic. The motivation of this work is to provide a string theory origin for anomalies, gaugings in parameter space, and the resulting decomposition. A key insight from string theory is that there are no truly free parameters; what appear as parameters in the effective QFTs engineered from string theory are actually the vacuum expectation values (vevs) of dynamical moduli fields. We aim to show that promoting these moduli to partially dynamical fields is a natural extension within string theory, where both the parameters and their associated topological local operators can be traced back to the presence of branes. Additionally, the domain walls that separate distinct universes in the decomposition can also be understood as being constructed from branes.

We illustrate our approach by examining QFTs derived from string theory and M-theory on manifolds with conical singularities, potentially with brane probes. This setup enables us to explore the geometric origin of QFT parameters across various dimensions. To investigate the anomalies and gauging in parameter space, we generalize the notion of symmetry theory \cite{Turaev:1992hq, Apruzzi:2021nmk, Freed:2022qnc}, defined in $(D+1)$-dimensions, for parameters. One essential property of this generalized symmetry theory for parameters, compared to the conventional ones, is there exist non-trivial topological local operators built from Euclidean branes. Thus, the symmetry theory enjoys multiple vacua as a topological field theory. As a result, the physical boundary theory is not a single relative theory \cite{Freed:2012bs}, but \emph{a class of} relative theories.

We separately consider symmetry theories for discrete and continuous parameters, and their derivation from string theory:
\begin{itemize}
     \item \textbf{For discrete parameters,} we derive symmetry theories by extending the top-down framework introduced in \cite{Apruzzi:2021nmk, GarciaEtxebarria:2024fuk} (see also \cite{Freed:2006yc, GarciaEtxebarria:2019caf} for earlier related discussions).
    \item \textbf{For continuous parameters,} we obtain symmetry theories by taking a topological limit of the kinetic term for moduli fields inherited from higher-form gauge fields in string theory. Notably, our results align with the $U(1)$ symmetry theory proposed in \cite{Brennan:2024fgj, Antinucci:2024zjp}, whose string theory origin, to our knowledge, has not been previously discussed in the literature \footnote{See also \cite{Apruzzi:2024htg, Bonetti:2024cjk, Cvetic:2024dzu} for alternative formulations of continuous symmetry theories from string theory.}.
\end{itemize}
In both cases, the gauging in the parameter space is manipulated as changing the topological boundary condition at the asymptotic boundary of the internal manifold in string theory and M-theory. Furthermore, the Chern-Simons couplings in string theory and M-theory under dimensional reduction descend to the anomalies in the parameter space, which have been discussed field-theoretically in \cite{Cordova:2019jnf, Cordova:2019uob}. After gauging the parameter space, the terms previously interpreted as anomalies indicate that the resulting QFT exhibits a higher-group structure, more precisely, a $D$-group structure in a $D$-dimensional QFT, reflecting the interplay between decomposition and other global symmetries. The topological operators, along with domain walls separating distinct universes, are realized through branes, generalizing the `branes at infinity' idea introduced in \cite{Apruzzi:2022rei, GarciaEtxebarria:2022vzq, Heckman:2022muc, Heckman:2022xgu}.


\section{\label{sec: symth}Symmetry Theory of Parameters from String Theory}
In this section, we present a general strategy for deriving symmetry theories for parameters from string theory. The resulting symmetry theory involves moduli fields and various boundary conditions for them. The Dirichlet boundary condition leads to a fixed profile for moduli fields, which correspond to parameters with fixed values or their promotion to background fields. Neumann boundary condition admits the moduli fields to be partially dynamical, constructing topological local operators and generating decomposition of the underlying effective QFT.

Consider a string theory on a $D$-dimensional QFT engineered from string theory on a $(10-D)$-dimensional internal manifold $X$ with a conical singularity (for M-theory a $(11-D)$-dimensional internal manifold):
\begin{equation}
    X=\text{Cone}(\partial X),
\end{equation}
with link $\partial X$ as the asymptotic boundary of $X$. One can parametrize the radial direction of the cone with coordinate $r$, so that $r=0$ is the tip of the cone supporting the singularity, while the asymptotic boundary $\partial X$ is located `at infinity' $r=\infty$.

The associated symmetry theory is a $(D+1)$-dimensional theory derived from the string compactification on the asymptotic boundary $\partial X$ of the internal geometry $X$ \cite{Apruzzi:2021nmk}. Schematically, the $q$-form gauge field $A_q$ for a $(q-1)$-form global symmetry in the effective QFT are derived from the reduction of a $p$-form NSNS or RR field (for M-theory $C_3$ field) on submanifold $\Sigma_{p-q} \subset X$:
\begin{equation}
    A_q \sim \int_{\Sigma_{p-q}}C_p.
\end{equation}

We generalize this construction to the case where $q=0$. That is to say, a $C_p$ gauge fields in string theory are dimensionally reduced on a $p$-cycle to a scalar modulus field $\varphi$ in $(D+1)$-dimensional spacetime, schematically expressed as $\varphi \sim \int_{\Sigma_p}C_p$. The modulus field $\varphi$ can be continuous or discrete, according to the submanifold $\Sigma_p$ to be torsional or not, respectively. 

\subsection{Symmetry theory for continuous parameters}
In this case, we want $\Sigma_p$ to be torsion-free. It is then easy to assume the internal manifold $X$ has only torsion-free submanifolds without loss of generality. The 10-dimensional kinetic term for $C_p$ can be dimensionally reduced on $\partial X$ to a $(D+1)$-dimensional Maxwell-type term for the modulus field $\varphi$:
\begin{equation}\label{eq: top trivial reduction of Maxwell}
   \int_{M_{D+1}\times \partial X}dC_p\wedge *_{10} dC_p \Rightarrow \frac{1}{2e^2}\int_{M_{D+1}}d\varphi \wedge *_{D+1} d\varphi,
\end{equation}
where $C_p$  and $\varphi$ are all $U(1)$ fields, and $*_{d}$ denotes the Hodge star operator in $d$ dimensions. 

\subsubsection{Half-magnetic formulation and topological limit}
We are interested in the topological limit of this theory to derive the symmetry theory for the QFT engineered on $M_D=\partial M_{D+1}$ associated with the string compactification on $X$. It is useful to reformulate the above Maxwell-type theory following \cite{Witten:1995gf}, 
\begin{equation}\label{eq: abelian duality}
\begin{split}
    &\int \mathcal{D\varphi} \exp\left[{\frac{i}{2e^2}\int_{M_{D+1}}d\varphi \wedge *_{D+1} d\varphi}\right]\\
    =&\int \mathcal{D}f_D\mathcal{D}\varphi \exp \left[\int_{M_{D+1}}\frac{ie^2}{2}f_{D}\wedge *_{D+1}f_{D}+d\varphi \wedge f_{D}\right],
\end{split}
\end{equation}
where $f_D$ is a arbitrary $D$-form (i.e. $\mathbb{R}$-valued) instead of a field-strength of any $U(1)$ gauge field. The equivalence of the second line to the first line of the above equation can be seen by noticing that $\varphi$ serves as a Lagrangian multiplier. Path integrating it over ends up with the condition $df_D=0$, implying the $f_D$ is now a field-strength of a $(D-1)$-form gauge field $dA_{D-1}=f_D$. The quantum theory is then 
\begin{equation}
    \int \mathcal{D}A_{D-1}\exp \left[\frac{i}{2\tilde{e}^2}\int_{M_{D+1}}dA_{D-1}\wedge *_{D+1}dA_{D-1}\right],
\end{equation}
with $\tilde{e}^2\equiv \frac{1}{e^2}$. This is exactly the magnetic dual description for the $\varphi$ theory (which can be regarded as the `electric' one) under abelian S-duality \cite{Witten:1995gf}. The second line of (\ref{eq: abelian duality}), in this sense, can be regarded as a `half-magnetic' formulation \footnote{We thank I. Garcia Etxebarria for suggesting this name}.

The half-magnetic formulation includes a topological term $d\varphi \wedge f_D$, so it likely admits a manifest topological limit to get the symmetry theory. In fact, the topological limit can be taken by noticing that the $\partial X$ is the asymptotic boundary of $X$ `at infinity' along the `radial' direction of the conical internal geometry $X$, which means it has an infinite volume. To be explicit, we parameterize this radial direction for the cone by coordinate $r$. The topological limit taken is $r\rightarrow \infty$. Recall that under compactification, the effective coupling constant is typically proportional to the inverse of the volume
\begin{equation}
    e^2 \sim \frac{1}{\text{vol}(\partial X)}\rightarrow 0.
\end{equation}
Now it is clear that under a large volume limit for $\partial X$, the metric-dependent term in the half-magnetic action degenerates, while the topological term persists \footnote{Similar limit was also discussed in the context of non-abelian continuous symmetries in \cite{Bonetti:2024cjk}.}. That is to say; we indeed take a topological limit for the $(D+1)$-dimensional theory, naturally realized as a symmetry theory
\begin{equation}\label{eq: symmetry theory for U(1) parameters}
    S_{D+1}=\int_{M_{D+1}}d\varphi \wedge f_D.
\end{equation}

Remarkably, since the modulus field $\varphi$ in this case is compact and $f_D$ is an unconstrained $D$-form, $S_{D+1}$ can be regarded as a theory of a $U(1)$ gauge field coupled to a $\mathbb{R}$ gauge field. This exactly recovers the $U(1)$ symmetry theory proposed in \cite{Brennan:2024fgj,Antinucci:2024zjp}.

\subsubsection{Topological operators and defects from (flux)branes}The symmetry theory has two classes of topological operators
\begin{equation}
\begin{split}
    &U_\alpha=\exp \left[i\alpha \oint_{M_D}f_D  \right], ~\alpha \in \mathbb{R}/\mathbb{Z}\cong U(1),\\ 
    &V_m=\exp \left[ im\varphi \right], ~m\in \mathbb{Z},
\end{split}
\end{equation}
where $U_\alpha$ are domain walls and $V_m$ are local operators. For QFTs constructed from string theory, it is believed that operators in the symmetry theory can be built via the topological sector of branes 'at infinity'. This has been discussed in the context of holography \cite{GarciaEtxebarria:2022vzq, Apruzzi:2022rei}, geometric engineering \cite{Heckman:2022muc}, and for general brane probes of singularities \cite{Heckman:2022xgu, Yu:2023nyn, Franco:2024mxa}. Following this general strategy, we now discuss the brane origin for $U_\alpha$ and $V_m$ operators in the symmetry theory for parameters. 

For the local operator $V_m$, notice that its associated $\varphi$ field comes from the dimensional reduction of the string theory higher gauge field $C_p$ integral on a non-torsional cycle $\Sigma$. It is easy to conclude that $V_m$ comes from the ordinary $(p-1)$-brane (i.e, D-brane or M-brane) electrically coupled to $C_p$, with its whole worldvolume defined on $\Sigma_p$. The topological sector of the brane is given by the Wess-Zumino term \cite{Douglas:1995bn} 
\begin{equation}\label{eq: wz term of ord brane}
    S_{\text{WZ}}^{\text{ord}}=\exp \left[im\int_{\Sigma_p}C_p+\cdots  \right],
\end{equation}
where $\cdots$ denotes the subleading terms irrelevant to our current discussion. $m$ labels the Dirac-quantized charge carried by the $(p-1)$-brane. 

The brane origin is more subtle for the domain wall operator $U_\alpha$. Its associated field $f_D$ is not a $U(1)$ gauge field, but an unconstrained $\mathbb{R}$-valued $D$-form field. As we discussed around the half-magnetic formulation (\ref{eq: abelian duality}) of the Maxwell-type theory, after integrating out the electric degrees of freedom $\varphi$, $f_D$ is interpreted as the field-strength for the dual magnetic $U(1)$ field $A_{D-1}$. It is related to the electric field-strength while S-duality $dA_{D-1}=f_D \sim *_{D+1}{d\varphi}$, but at the level of the topological limit where both $\varphi$ and $f_D$ are present, it is more proper to treat $f_D$ as an unconstrained \emph{magnetic flux}, instead of a field-strength. It is then natural to propose its string theory origin as the magnetic dual flux $G_{9-p}$ for the electric gauge field $C_p$: $G_{9-p}\sim *_{10} dC_p$. The brane coupled to this flux $G_{9-p}$ is the $(8-p)$-fluxbrane \cite{Gutperle:2001mb, Emparan:2001gm}, whose Wess-Zumino term enjoys a similar form as the ordinary brane
\begin{equation}\label{eq: wz term of fluxbrane}
    S_{\text{WZ}}^{\text{flux}}=\exp \left[ i\alpha \int_{M_D\times \Sigma_{9-D-p}}G_{9-p}+\cdots \right]
\end{equation}
where $\cdots$ denotes the subleading terms irrelevant to our current discussion. The worldvolume of this flux brane is defined on $M_D\times \Sigma_{9-D-p}$, where $\Sigma_{9-D-p}$ is the dual cycle intersecting with $\Sigma_p$ within the $(9-D)$-dimensional asymptotic boundary manifold $\partial X$. 

\subsubsection{Topological boundary conditions}
Having derived a $(D+1)$-dimensional symmetry theory for parameters, one asks what topological boundary conditions are allowed and their implication to the resulting $D$-dimensional QFT. In the string theory context, this amounts to picking a boundary condition at $r=\infty$ and specifying the parameter space of the QFT engineered at the local singularity at $r=0$.

The conventional setup for QFT with fixed parameters or parameters promoted to background fields corresponds to a Dirichlet condition for $\varphi$
\begin{equation}\label{eq: Dirichlet condition for U(1) SymTFT}
   \varphi|=\lambda.
\end{equation}
This includes the case when $\lambda$ is a fixed parameter and when $\lambda=\lambda(x)$ is a background field with a fixed profile. Local operators $V_m$ are frozen/trivialized by this boundary condition for all $m$. This boundary condition implies a  $U(1)$ `$(-1)$-form symmetry', generated by the topological operator $U_\alpha$, while $\varphi$ is the 0-form background field. Recall that $U_\alpha$ are built from $(8-p)$-fluxbranes; this boundary condition thus aligns with the result in \cite{Cvetic:2023plv} where authors proposed fluxbranes constructing the $U(1)$ symmetry operators \footnote{Another proposal for branes behind $U(1)$ symmetry operators was discussed in \cite{Bergman:2024aly}, where the Wess-Zumino term (\ref{eq: wz term of fluxbrane}) was identified as the topological sector of non-BPS branes.}.

We remark that changing the boundary condition field profile for $\varphi$ does not necessarily preserve the \emph{local} information of the QFT. In contrast, in the conventional symmetry theory, the topological boundary only specifies the global structure. For example, consider $\lambda$ as one of the coupling constants for the theory with action $S_\lambda[\phi]$, where $\phi$ denotes various fields in the theory. Stacking a $U_\alpha$ operator onto the topological boundary will change the theory 
\begin{equation}
\begin{split}
    &S_\lambda[\phi] = S_0[\phi]+ \int d^Dx~\lambda O(x)\\
    \rightarrow &S_{\lambda+\alpha}[\phi]=S_0[\phi]+\int d^Dx~(\lambda+\alpha) O(x).
\end{split}
\end{equation}
This is a deformation for the theory $S_{\lambda}[\phi]\rightarrow S_{\lambda+\alpha}[\phi]$, which in general will change the local information. That is to say, the non-topological boundary theory for the symmetry theory is not a single theory with specific local information, but a family of QFTs building a deformation class.

We are now interested in promoting the parameter $\lambda$ as a partially dynamical field. This translates in the following boundary condition (\ref{eq: Dirichlet condition for U(1) SymTFT}) to 
\begin{equation}\label{eq: Z_N condition for U(1) SymTFT}
    N\varphi|=\lambda.
\end{equation}
This leads to an equivalence of $V_m\sim V_{m+N}$. That is to say, local operators $V_n$ with $n\in \mathbb{Z}/N\mathbb{Z}\cong \mathbb{Z}_N$ can still be dynamical along the topological boundary. These non-trivial topological local operators generate a $\mathbb{Z}_N$ $(D-1)$-form symmetry for the resulting $D$-dimensional QFT.  

This $(D-1)$-form symmetry prevents a fixed valued or background field profile for the parameter $\lambda$ and its associated modulus field $\varphi$. The partition function of the $D$-dimensional theory is written as 
\begin{equation}\label{eq: decomposition partition function}
\begin{split}
    Z&=\sum_{n=1}^{N}\int \mathcal{D}\phi~\exp [iS[\phi]] \exp[i\frac{n\lambda}{N}\int d^Dx O(x)]
\end{split}
\end{equation}
which is a sum over partition functions of N theories labeled by $Z_n=\int \mathcal{D}O~\exp [iS] \exp[i\frac{n\lambda}{N}\int d^Dx O(x)]$. In other words, the theory decomposes into $N$ universes \cite{Tanizaki:2019rbk, Komargodski:2020mxz, Sharpe:2022ene}.

A key difference between universes and conventional superselection sectors is that though they are both separated by domain walls, those separating universes have infinite tension. This aligns with the $(D-1)$-form symmetry interpretation of decomposition, where these domain walls are precisely the $(D-1)$-dimensional charged defects. These defects come from a subset of $U_\alpha$ operators labeled by $U_n$ with $n\in \mathbb{Z}_N$ that can terminate on the topological boundary.

Recall that the local operators $V_m$ and domain walls $U_\alpha$ are originated from ordinary branes (\ref{eq: wz term of ord brane}) and fluxbranes (\ref{eq: wz term of fluxbrane}). We can then embed the above discussion in the string theory, and summarize the brane interpretation for the gauging in the parameter space:
\begin{itemize}
    \item Euclidean branes lying at the asymptotic boundary $r=\infty$ are topological local operators generating the $(D-1)$-form symmetry, decomposing the $D$-dimensional QFT.
    \item Fluxbranes attaching from the singularity $r=0$ to the asymptotic boundary $r=\infty$ are infinitely heavy domain walls for the $D$-dimensional QFT, separating different universes.
\end{itemize}

\subsection{Symmetry theory for discrete parameters}\label{sec: discrete symtft}

In this case, we are interested in $\Sigma_p$ wrapped by a $p$-(spacetime)dimensional brane to be torsional. The dimensional reduction of the higher gauge field $C_p$ in string theory now leads to a discrete parameter. Roughly speaking, the symmetry theory in this case also comes from the Maxwell-type action for the $C_p$ field. However, compared to the topologically trivial (i.e., non-torsional) reduction for the continuous parameter in (\ref{eq: top trivial reduction of Maxwell}), the dimensional reduction for the discrete case needs to be treated with carefulness. Let us investigate the 10D string theory case, after which the 11D M-theory case will work similarly. 

According to \cite{Belov:2006jd, Freed:2006ya, GarciaEtxebarria:2024fuk, Yu:2023nyn}, for topologically non-trivial background, the 10D Maxwell-type action in string theory should be formulated as a boundary theory of an 11D topological theory in terms of the differential cohomology \cite{Freed:2006yc}\footnote{See also, e.g., \cite{Heckman:2017uxe, Bah:2020jas, Bah:2020uev, Apruzzi:2023uma, Lawrie:2023tdz} on derivation of discrete symmetry theories from 11D formulism of string theory.}
\begin{equation}
     \int_{M_{D+1}\times \partial X}dC_p\wedge *_{10} dC_p \rightarrow \int_{N_{D+2}\times \partial X}\breve{G}_{p+2} \star \breve{G}_{10-p}.
\end{equation}
In the above expression, $\breve{G}_{p+2}$ is the differential cohomology uplift of the 11D topological gauge field $\tilde{C}_{p+1}$, whose gauge transformation parameter corresponds to the $C_p$ gauge field in 10D string theory. $N_{D+2}$ is an auxiliary manifold whose boundary $\partial N_{D+2}=M_{D+1}$ is where the symmetry theory will live. 

The dimensional reduction of $\breve{G}_{p+2}$ on torsional cycle $\Sigma_p$ leads to a differential 1-cochain $\breve{a}_1 \sim \int_{\Sigma_p}\breve{G}_{p+2}$ (and similarly for the reduction $\breve{G}_{10-p}$). We can then obtain the dimensional reduced action 
\begin{equation}
   \int_{N_{D+2}\times \partial X}\breve{G}_{p+2} \star \breve{G}_{10-p}\Rightarrow \Omega \int_{N_{D+2}}\breve{a}_1 \star \breve{b}_{D+1},
\end{equation}
where $\Omega$ is determined by the linking number of $\Sigma_p$ and its linking cycle.
Assuming the external spacetime $N_{D+2}$ as well as its boundary $M_{D+1}$ have no torsional closed-submanifold, the above action reduces to a $(D+1)$-dimensional topological field theory which can be written as ordinary cochains or differential forms \cite{GarciaEtxebarria:2024fuk}. To be explicit, we consider $\Sigma_p$ being dual to a $\mathbb{Z}_N$ cohomology generator, the resulting action reads
\begin{equation}\label{eq: zn gauge theory}
    S_{D+1}=N\int_{M_{D+1}}a_0 \wedge db_D.
\end{equation}
This is a generalization of $(D+1)$-dimensional $\mathbb{Z}_N$ gauge theory to a 0-form field $a_0$.

\subsubsection{Topological operators and defects from branes}
The above generalized $\mathbb{Z}_N$ gauge theory has the equations of motion $Nda_0=Ndb_D=0$. This leads to the following finite set of topological operators 
\begin{equation}
\begin{split}
    &U_m=\exp \left[im \oint_{M_D}b_D  \right],\\ 
    &V_m=\exp \left[ ima_0 \right], ~m=0,\cdots, N-1
\end{split}
\end{equation}

The brane origin for $V_m$ operators is similar to the case of continuous parameters. Recall that $a_0$ is derived from the reduction of the (differential cohomology uplift of) higher-form gauge field $C_p$, so its associated operator $V_m$ is built from the topological limit of the $(p-1)$-brane wrapping on torsional cycle $\Sigma_p$. While for $U_m$ operators, the underlying branes are no longer fluxbranes as in the continuous parameter case but ordinary branes wrapping the torsional linking cycle of $\Sigma_p$ in $\partial X$. Furthermore, this ordinary brane is the magnetic dual $(7-p)$-brane ($(8-p)$-brane respectively in M-theory) of the $(p-1)$-brane \cite{Heckman:2022muc}.

\subsubsection{Topological boundary conditions}
The discrete parameter for the $D$-dimensional QFT engineered on $r=0$ is realized by the electric boundary condition at $r=\infty$ for the symmetry theory (\ref{eq: zn gauge theory})
\begin{equation}
    a_0|=\chi.
\end{equation}
This is a Dirichlet condition for $a_0$, where $\chi$ is a fixed parameter taking values in $\{0,\cdots N-1\}$. $\chi$ thus plays the role of a background gauge field for a $\mathbb{Z}_N$ `$(-1)$-form symmetry'. The symmetry generators are $U_m$ operators, which are spacetime-filling from the $D$-dimensional QFT perspective.

As discussed in the continuous parameter case, before specifying the boundary condition at $r=\infty$, the theory living at $r=0$ is not a conventional relative QFT with certain local information, but a family of theories labeled by different values of $\chi$.  Changing the value of $\chi$ in general will change the local information, e.g., the rank of the gauge group.

Promoting the parameter $\chi$ as a dynamical field is less ambiguous than the continuous case due to the discrete nature of the $a_0$ field. Recall that in the continuous case, $\varphi$ field is a $U(1)$ compact scalar, so performing the promotion with or without introducing the kinetic terms is possible. What we choose is to limit $\varphi$ to be flat so that it does not propagate but only involves summing over discrete topologies, which we interpret as gauging a finite subgroup of $U(1)$ `$(-1)$-form symmetry'. When it comes to the discrete case, it is easy to see $a_0$ itself is a gapped $\mathbb{Z}_N$ gauge field; there is no propagating kinetic term one can introduce for it. A magnetic boundary condition realizes the promotion of $\chi$
\begin{equation}
    b_D|=B_D.
\end{equation}
This Dirichlet boundary condition of $b_D$ freezes the fluctuation of $U_m$ operators while keeping the $V_m$ operators to be dynamical. These local operators $V_m$ then build a set of projection operators and thus decompose the theory. The partition function of the theory is similar to (\ref{eq: decomposition partition function}),
\begin{equation}\label{eq: decomposition discrete partition function}
\begin{split}
    Z&=\sum_{\chi=0}^{N-1}\int \mathcal{D}\phi~\exp [iS_\chi[\phi]]
\end{split}
\end{equation}
where $S_\chi[\phi]$ is the action for theory labeled by the discrete parameter $\chi$. This leads to $N$ universes, implying a $\mathbb{Z}_N$ $(D-1)$-form symmetry generated by $V_m$.

Recall that the local operators $V_m$ and domain walls $U_m$ are originated from ordinary branes and its dual magnetic branes. We can then embed the above discussion in the string theory, and summarize the brane interpretation for the gauging in the parameter space:
\begin{itemize}
    \item Euclidean $(p-1)$-branes lying at $r=\infty$ are topological local operators generating the $(D-1)$-form symmetry, decomposing the $D$-dimensional QFT.
    \item Magnetic dual $(7-p)$-branes attaching from singularity $r=0$ to $r=\infty$ are domain walls with infinite tension, separating different universes.
\end{itemize}

\section{Examples}\label{sec: example}

\subsection{Continuous parameter case: Modified instanton sum in $4D$ $\mathcal{N}=4$ SYM}
Let us first consider gauging in the continuous parameter space. One of the most celebrated stages to discuss the various behaviors of parameters is the $\theta$-angle in the 4D $SU(K)$ gauge theory. Notably, in \cite{Cordova:2019uob}, it was realized that there is a mixed anomaly between the $2\pi$-shift of $\theta$ and the 1-form $\mathbb{Z}_K$ symmetry, and then in \cite{Tanizaki:2019rbk} the modified instanton sum labeled by $\theta$ is discussed. What we will do in this section is discuss the $\mathcal{N}=4$ supersymmetric version of the $SU(K)$ theory, and investigate the symmetry theory of the $\theta$-angle from branes, providing a top-down perspective that naturally unifies previous field-theoretic results.

Consider $K$ D3-branes probing the transverse flat space $\mathbb{C}^3$ in Type IIB string theory. The low-energy effective theory on their worldvolume $M_4$ is 4D $\mathcal{N}=4$ super Yang-Mills theory with Lie algebra $\mathfrak{su}(K)$. The $\theta$-angle related to the instanton sector $\int_{M_4}\theta F_2\wedge F_2$ of the gauge theory is typically given by the vev of the IIB axion field $C_0$. Instead of fixing its vev, we follow our general strategy in Section II and derive its symmetry theory in $M_5=M_4\times \mathbb{R}_+$ by dimensionally reducing the kinetic term for $C_0$ on the asymptotic boundary $\partial \mathbb{C}^3=S^5$ and taking the topological limit. The resulting symmetry theory term reads
\begin{equation}
    \int_{M_5}d C_0 \wedge f_4,
\end{equation}
The conventional gauge theory with a certain $\theta$-angle is reproduced by picking a Dirichlet boundary condition $C_0|=\theta$ at the asymptotic boundary $r=\infty$. Branes behind this $U(1)$ `$(-1)$-form symmetry' are 8-fluxbranes wrapping on $S^5$ with worlvolume flux $\int_{S_5}G_9=f_4$, where $G_9$ is the magnetic flux dual to the electric field $C_0$ coupled to D$(-1)$-brane, i.e. D-instanton.

Notice that so far, we haven't specified the global form of the gauge group. To determine the global structure, one considers the symmetry theory for 1-form symmetry, schematically obtained from the 10D Chern-Simons term \cite{Witten:1998wy}. Usually, this is done without turning on axion $C_0$ (namely, no 7-brane effect). However, here we are also interested in the symmetry theory for the $\theta$-angle, so we turn on the $C_0$ field and use the complete physical 3-form field-strength $F_3=dC_2-C_0\wedge dB_2$ \cite{Bergman:2022otk} (see also \cite{GarciaEtxebarria:2024jfv}):
\begin{equation}
    \int_{M_{5}\times S^5}F_5\wedge B_2\wedge F_3 \Rightarrow N \int_{M_5} B_2\wedge dC_2 - N\int_{M_5}dC_0\wedge  B_2 \wedge B_2\,.
\end{equation}
We thus obtain the symmetry theory 
\begin{equation}\label{eq: sym theory for SYM}
    S_5=\int_{M_5}dC_0\wedge f_4 + N \int_{M_5} b_2\wedge dC_2 - N\int_{M_5}dC_0\wedge  B_2 \wedge B_2.
\end{equation}

Picking the Dirichlet condition for $C_0$ and $B_2$, we realized the conventional $SU(K)$ gauge theory with certain $\theta$. Furthermore, the cubic term in (\ref{eq: sym theory for SYM}) reduces to an invertible TFT
\begin{equation}
    -N\int_{M_5}d\theta \wedge B_2 \wedge B_2,
\end{equation}
reproducing the mixed anomaly between the $\theta$ parameter and the $\mathbb{Z}_N$ 1-form symmetry discovered in \cite{Cordova:2019uob}. 

Let us now consider the boundary condition (\ref{eq: Z_N condition for U(1) SymTFT}) for $C_0$: $NC_0|=\theta$. Under this boundary conditions, a $\mathbb{Z}_N$ sector of D-instantons are admitted to lying along the asymptotic boundary $S^5$, i.e. `at infinity' from the D3-branes worldvolume. This provides $N$ topological local operators $\exp [i n C_0 ], n=0,1,\cdots N-1$ in the resulting 4D gauge theory, decomposing the theory into $N$ universes, labeled by a discrete parameter $\frac{n}{N}\theta$. From a 4D perspective, this is exactly performing a $\mathbb{Z}_N$ modified instanton sum discussed in \cite{Tanizaki:2019rbk}. The quantum 3-form $\mathbb{Z}_N$ symmetry generated by D-instantons have charged defects, which are just magnetic dual 8-fluxbranes wrapping the cone over $S^5$, namely attaching between the D3-branes at $r=0$ to the infinity at $r=\infty$. The non-compactness of these wrapped branes aligns with the fact that they build infinitely heavy domain walls separating different universes in the 4D theory.

Moreover, this D-instanton `at infinity' setup produces not only a modified instanton sum, but also a higher-group structure. The equation of motion for $C_0$ tells us 
\begin{equation}
    df_4=2N B_2 dB_2.
\end{equation}
Under the boundary condition where a $\mathbb{Z}_N$ 3-form symmetry is present, the boundary profile of $f_4$ will serve as the background field for this 3-form symmetry. However, the above equation shows $f_4$ is flat only up to the twist by the $\mathbb{Z}_K$ 1-form symmetry background $B_2$. This is thus a 4-group structure $\mathcal{G}^{[4]}$ schematically expressed as 
\begin{equation}
\mathcal{G}^{[4]}=\mathbb{Z}_N^{(3)}\tilde{\times} \mathbb{Z}_K^{(1)},
\end{equation}
which was discussed in the non-supersymmetry case in \cite{Tanizaki:2019rbk}. In the above equation we use $\tilde{\times}$ to denote it is not a direct product.

\subsection{\label{sec: Exa2}Discrete parameter case: Gauging gauge ranks in 3D ABJ(M) theories}

Having discussed how to gauge in the continuous parameter space and perform the modified instanton sum from a top-down perspective, we now move to the case of gauging discrete parameters. We set up our stage in 3D and QFTs constructed from M-theory.  Unlike the 4D examples whose field-theoretic result is known in closely similar theories, the gauging manipulation, its resulting decomposition, and higher-group structures we will discuss in this section, to the best of our knowledge, have not appeared in the literature.

An infinite family of 3D QFTs can be constructed on M2-branes probing Calabi-Yau 4-fold singularities. The most celebrated example is the ABJ(M) theory with the singular Calabi-Yau as an abelian orbifold, i.e., $X=\mathbb{C}^4/\mathbb{Z}_k$. Given $N$ M2-brane probes, the 3D QFT living on the worldvolume of M2-branes is a holographic SCFT \cite{Aharony:2008ug}, with a Chern-Simons-matter theory description $U(N)_k\times U(N)_{-k}$. In addition to the regular M2-branes, it is also possible to consider fractional M2-branes as M5-branes wrapping torsional 3-cycles $\gamma_3$ \cite{Aharony:2008gk}. $M$ such fractional branes modify the rank of the gauge group, leading to a $U(N)_k\times U(N+M)_{-k}.$ theory. In particular, the $M$ can only pick values from $\{ 0, 1, \cdots k-1 \}$.

Let us now treat the modified gauge rank $M$ as a discrete parameter. Schematically, this discrete parameter, counting the number of fractional M2-branes, comes from the dimensional reduction of the $C_3$ gauge field on $\gamma_3$ in M-theory.  Following our general strategy in Section \ref{sec: discrete symtft}, its associated symmetry theory can be derived from M-theory action on a 12D manifold $M_{12}=N_5\times \partial X$ via the dimensional reduction on $\partial X=S^7/\mathbb{Z}_7$:
\begin{equation}\label{eq: BF symtft for ABJM}
	\int_{N_5\times \partial X}\breve{G}_5\star \breve{G}_8\Rightarrow 
	k\int_{M_4}b_0 \cup dc_3,
\end{equation}
where  $M_4=\partial N_5$ is the 4-manifold bulk for the symmetry theory.

The number of the modified gauge rank $M$ is reproduced via the boundary condition $b_0|=M$ at $r=\infty$. From a generalized symmetry point of view, this specifies a $\mathbb{Z}_k$ `$(-1)$-form symmetry', whose symmetry generators are $\exp \left[ im\int_{M_3}{c_3}\right]$, whose brane origin is precisely the M5-brane wrapping on $\gamma_3$, spacetime-filling in the 3D QFT living on $M_3$.

Gauging in this discrete parameter space is realized via changing the boundary condition to $c_3|=C_3$ with $b_0$ free. This Neumann condition for $b_0$ admits the M2-branes wrapping on $\gamma'_3$, i.e., the linking torsional cycle of $\gamma_3$, to lie along the asymptotic boundary at $r=\infty$. This wrapped M2-brane is now a Euclidean brane `at infinity', giving rise to topological local operators $\exp \left[ imb_0 \right]$. These local operators generate a $\mathbb{Z}_k$ 2-form symmetry for the resulting 3D QFT. Due to these operators, the partition function of the 3D QFT reads
\begin{equation}
    Z=\sum_{i=0}^{k-1}Z_{i}^{\text{ABJ(M)}},
\end{equation}
i.e., the theory decomposes into a disjoint union of $k$ universes, each universe given by an ABJ(M) theory. The 2D domain walls separating different ABJ(M) universes are engineered from M5-branes wrapping $\text{Cone}(\gamma_3)$, attaching the probing M2-branes at $r=0$ to the asymptotic boundary $r=\infty$.

Moreover, this 3D decomposition not only implies a $2$-form symmetry but also interplays with other global symmetries and leads to a higher-group structure. Performing the dimensional reduction for the Chern-Simons theory $C_3\wedge dC_3 \wedge dC_3$ on $\partial X$, one ends up with a cubic interaction in $M_4$, schematically:
\begin{equation}
    -\Omega \int_{M_4}b_0\wedge b_2\wedge b_2,
\end{equation}
where $b_2$ is the background field for the 1-form symmetry of the ABJ(M) theory, and $\Omega$ is the coefficient determined by the geometric data, whose specific result is irrelevant to our current discussion. In \cite{vanBeest:2022fss}, $b_0$ is treated as a fixed parameter, and $\Omega b_0$ is then regarded as capturing the self 't Hooft anomaly for the 1-form symmetry. However, in our current context, $b_0$ also fluctuates in the symmetry theory bulk $M_4$. Together with (\ref{eq: BF symtft for ABJM}), the symmetry theory terms relevant to $b_0$ read
\begin{equation}\label{eq: full symtft in ABJM}
    k\int_{M_4}b_0 \cup dc_3-\Omega \int_{M_4}b_0\wedge b_2\wedge b_2.
\end{equation}
Now, we consider picking the Dirichlet boundary condition for both $c_3$ and $b_2$, which leads to 2-form symmetry (i.e., decomposition) and also 1-form symmetry with background field $B_2=b_2|$. 
The equation of motion for $b_0$ in (\ref{eq: full symtft in ABJM}) tells us
\begin{equation}
    kdC_3=\Omega B_2\wedge B_2.
\end{equation}
This non-closed condition $kdC_3\neq 0$ implies the fact that there is a higher-group symmetry \cite{Cordova:2018cvg}, more precisely a 3-group $\mathcal{G}^{[3]}$, as a extension of 1-form symmetry $G^{(1)}$ by the decomposition 2-form $\mathbb{Z}_k^{(2)}$ symmetry 
\begin{equation}
    \mathcal{G}^{[3]}=\mathbb{Z}_k^{(2)} \tilde{\times}G^{(1)}.
\end{equation}

\section{Discussion}
We discuss a generalized notion of symmetry theory for the parameter space in QFTs, and investigate their gauging and anomalies. We provide a natural top-down treatment so that both symmetry theory and various operators, have a string theory interpretation. We illustrate with examples that one general consequence of gauging in parameter space is the resulting decomposition mixes with other global symmetries to build a $D$-group in $D$-dimensional theories.

Our work suggests various natural directions for future investigation. One is to consider turning on a discrete parameter $n$ for 5D QFTs engineered from M-theory on Calabi-Yau 3-folds, e.g. $\mathbb{C}^3/\mathbb{Z}_3$, with $n$ M5-branes wrapping on torsional 1-cycles (case $n=0$ is the well-known 5D $E_0$ SCFT \cite{Seiberg:1996bd}.). With nonvanishing $n$, it is suspected that the 5D QFT can be gapped \footnote{Ethan Torres, private communication.}, thus gauging this discrete parameter might lead to an interesting decomposition where different universes can be either gapless or gapped. Another direction is to investigate how the parameter space can be interpreted as not just `(-1)-form symmetry', but even lower-form symmetry proposed in \cite{Heckman:2024oot}. One potentially useful example is to consider the Roman mass in IIA. Regarding it as a `(-1)-form symmetry' in gravity theory, then from the dual 3D QFT point of view \cite{Gaiotto:2009mv}, it would correspond to a `(-2)-form symmetry'. Investigating the gauging of this symmetry and the potential quantum $D$-form symmetry would be interesting. Finally, the decomposition from gauging the parameter is reminiscent of the ensemble averaging in the context of lower-dimensional AdS/CFT, in the sense that the partition function involves multiple QFTs. It was proposed in \cite{Heckman:2021vzx} that ensemble averaging can be mimicked from a top-down perspective. The philosophy in this paper is similar to that of summing/averaging over multiple QFTs due to the dynamic nature of the modulus field in string theory. It would be interesting to investigate the potential connection between gauging in parameter space and the ensemble averaging holography.

\begin{acknowledgments}
The author thanks E. Sharpe and Y. Zheng for their collaboration at an early stage of this project and for carefully reading the manuscript. The author thanks F. Apruzzi, I. Bah, T. D. Brennan, I. Garcia Etxebarria, J. J. Heckman, H. T. Lam, D. Rodriguez-Gomez, E. Sharpe, Z. Sun, E. Torres, and Y. Zheng for their valuable discussions. The author would like to thank the SCGP workshop ``Applications of Generalized Symmetries and Topological Defects to Quantum Matter", Simons Collaboration on Global Categorical Symmetries Annual Meeting 2024, Fudan Center for Field Theory and Particle Physics, Shanghai Institute for Mathematics and Interdisciplinary Sciences, and Tsinghua Yau Mathematical Sciences Center for their hospitality during part of this work. The author is supported by the NSF grant PHY-2310588. \textbf{Note added:} Top-down derivation for $U(1)$ symmetry theory is also covered in \cite{GarciaEtxebarria:toappear} which appears simultaneously with our work. We would like to thank them for agreeing to coordinate
the submission.

\end{acknowledgments}

\appendix


\bibliographystyle{unsrt}
\bibliography{parameter}

\begin{thebibliography}{10}

\bibitem{Seiberg:1994bp}
Nathan Seiberg.
\newblock {The Power of holomorphy: Exact results in 4-D SUSY field theories}.
\newblock In {\em {Particles, Strings, and Cosmology (PASCOS 94)}}, pages 0357--369, 5 1994.

\bibitem{Hellerman:2006zs}
Simeon Hellerman, Andre Henriques, Tony Pantev, Eric Sharpe, and Matt Ando.
\newblock {Cluster decomposition, T-duality, and gerby CFT's}.
\newblock {\em Adv. Theor. Math. Phys.}, 11(5):751--818, 2007.

\bibitem{Sharpe:2022ene}
Eric Sharpe.
\newblock {An introduction to decomposition}.
\newblock 4 2022.

\bibitem{Tanizaki:2019rbk}
Yuya Tanizaki and Mithat \"Unsal.
\newblock {Modified instanton sum in QCD and higher-groups}.
\newblock {\em JHEP}, 03:123, 2020.

\bibitem{Komargodski:2020mxz}
Zohar Komargodski, Kantaro Ohmori, Konstantinos Roumpedakis, and Sahand Seifnashri.
\newblock {Symmetries and strings of adjoint QCD$_{2}$}.
\newblock {\em JHEP}, 03:103, 2021.

\bibitem{Gaiotto:2014kfa}
Davide Gaiotto, Anton Kapustin, Nathan Seiberg, and Brian Willett.
\newblock {Generalized Global Symmetries}.
\newblock {\em JHEP}, 02:172, 2015.

\bibitem{Seiberg:2010qd}
Nathan Seiberg.
\newblock {Modifying the Sum Over Topological Sectors and Constraints on Supergravity}.
\newblock {\em JHEP}, 07:070, 2010.

\bibitem{Cordova:2019jnf}
Clay C{\' o}rdova, Daniel~S. Freed, Ho~Tat Lam, and Nathan Seiberg.
\newblock {Anomalies in the Space of Coupling Constants and Their Dynamical Applications I}.
\newblock {\em SciPost Phys.}, 8(1):001, 2020.

\bibitem{Cordova:2019uob}
Clay C{\' o}rdova, Daniel~S. Freed, Ho~Tat Lam, and Nathan Seiberg.
\newblock {Anomalies in the Space of Coupling Constants and Their Dynamical Applications II}.
\newblock {\em SciPost Phys.}, 8(1):002, 2020.

\bibitem{Sharpe:2019ddn}
Eric Sharpe.
\newblock {Undoing decomposition}.
\newblock {\em Int. J. Mod. Phys. A}, 34(35):1950233, 2020.

\bibitem{Vandermeulen:2022edk}
Thomas Vandermeulen.
\newblock {Lower-Form Symmetries}.
\newblock 11 2022.

\bibitem{Aloni:2024jpb}
Daniel Aloni, Eduardo Garc\'\i{}a-Valdecasas, Matthew Reece, and Motoo Suzuki.
\newblock {Spontaneously broken (-1)-form U(1) symmetries}.
\newblock {\em SciPost Phys.}, 17(2):031, 2024.

\bibitem{Brennan:2024tlw}
T.~Daniel Brennan.
\newblock {Constraints on symmetry-preserving gapped phases from coupling constant anomalies}.
\newblock {\em Phys. Rev. D}, 110(4):L041701, 2024.

\bibitem{Turaev:1992hq}
V.~G. Turaev and O.~Y. Viro.
\newblock {State sum invariants of 3 manifolds and quantum 6j symbols}.
\newblock {\em Topology}, 31:865--902, 1992.

\bibitem{Apruzzi:2021nmk}
Fabio Apruzzi, Federico Bonetti, I.~Garc\'\i{}a~Etxebarria, Saghar~S. Hosseini, and Sakura Schafer-Nameki.
\newblock {Symmetry TFTs from String Theory}.
\newblock 12 2021.

\bibitem{Freed:2022qnc}
Daniel~S. Freed, Gregory~W. Moore, and Constantin Teleman.
\newblock {Topological symmetry in quantum field theory}.
\newblock 9 2022.

\bibitem{Freed:2012bs}
Daniel~S. Freed and Constantin Teleman.
\newblock {Relative quantum field theory}.
\newblock {\em Commun. Math. Phys.}, 326:459--476, 2014.

\bibitem{GarciaEtxebarria:2024fuk}
I\~naki Garc\'\i{}a~Etxebarria and Saghar~S. Hosseini.
\newblock {Some aspects of symmetry descent}.
\newblock 4 2024.

\bibitem{Freed:2006yc}
Daniel~S. Freed, Gregory~W. Moore, and Graeme Segal.
\newblock {Heisenberg Groups and Noncommutative Fluxes}.
\newblock {\em Annals Phys.}, 322:236--285, 2007.

\bibitem{GarciaEtxebarria:2019caf}
I.~Garcia~Etxebarria, Ben Heidenreich, and Diego Regalado.
\newblock {IIB flux non-commutativity and the global structure of field theories}.
\newblock {\em JHEP}, 10:169, 2019.

\bibitem{Brennan:2024fgj}
T.~Daniel Brennan and Zhengdi Sun.
\newblock {A SymTFT for Continuous Symmetries}.
\newblock 1 2024.

\bibitem{Antinucci:2024zjp}
Andrea Antinucci and Francesco Benini.
\newblock {Anomalies and gauging of U(1) symmetries}.
\newblock 1 2024.

\bibitem{Note1}
See also \cite {Apruzzi:2024htg, Bonetti:2024cjk, Cvetic:2024dzu} for alternative formulations of continuous symmetry theories from string theory.

\bibitem{Apruzzi:2022rei}
Fabio Apruzzi, Ibrahima Bah, Federico Bonetti, and Sakura Schafer-Nameki.
\newblock {Noninvertible Symmetries from Holography and Branes}.
\newblock {\em Phys. Rev. Lett.}, 130(12):121601, 2023.

\bibitem{GarciaEtxebarria:2022vzq}
I.~Garcia~Etxebarria.
\newblock {Branes and Non-Invertible Symmetries}.
\newblock {\em Fortsch. Phys.}, 70(11):2200154, 2022.

\bibitem{Heckman:2022muc}
Jonathan~J. Heckman, Max H\"ubner, Ethan Torres, and Hao~Y. Zhang.
\newblock {The Branes Behind Generalized Symmetry Operators}.
\newblock 9 2022.

\bibitem{Heckman:2022xgu}
Jonathan~J. Heckman, Max Hubner, Ethan Torres, Xingyang Yu, and Hao~Y. Zhang.
\newblock {Top down approach to topological duality defects}.
\newblock {\em Phys. Rev. D}, 108(4):046015, 2023.

\bibitem{Witten:1995gf}
Edward Witten.
\newblock {On S duality in Abelian gauge theory}.
\newblock {\em Selecta Math.}, 1:383, 1995.

\bibitem{Note2}
We thank I. Garcia Etxebarria for suggesting this name.

\bibitem{Note3}
Similar limit was also discussed in the context of non-abelian continuous symmetries in \cite {Bonetti:2024cjk}.

\bibitem{Yu:2023nyn}
Xingyang Yu.
\newblock {Non-invertible Symmetries in 2D from Type IIB String Theory}.
\newblock 10 2023.

\bibitem{Franco:2024mxa}
Sebastian Franco and Xingyang Yu.
\newblock {Generalized symmetries in 2D from string theory: SymTFTs, intrinsic relativeness, and anomalies of non-invertible symmetries}.
\newblock {\em JHEP}, 11:004, 2024.

\bibitem{Douglas:1995bn}
Michael~R. Douglas.
\newblock {Branes within branes}.
\newblock {\em NATO Sci. Ser. C}, 520:267--275, 1999.

\bibitem{Gutperle:2001mb}
Michael Gutperle and Andrew Strominger.
\newblock {Fluxbranes in string theory}.
\newblock {\em JHEP}, 06:035, 2001.

\bibitem{Emparan:2001gm}
Roberto Emparan and Michael Gutperle.
\newblock {From p-branes to fluxbranes and back}.
\newblock {\em JHEP}, 12:023, 2001.

\bibitem{Cvetic:2023plv}
Mirjam Cveti\v{c}, Jonathan~J. Heckman, Max H\"ubner, and Ethan Torres.
\newblock {Fluxbranes, Generalized Symmetries, and Verlinde's Metastable Monopole}.
\newblock 5 2023.

\bibitem{Note4}
Another proposal for branes behind $U(1)$ symmetry operators was discussed in \cite {Bergman:2024aly}, where the Wess-Zumino term (\ref {eq: wz term of fluxbrane}) was identified as the topological sector of non-BPS branes.

\bibitem{Belov:2006jd}
Dmitriy Belov and Gregory~W. Moore.
\newblock {Holographic Action for the Self-Dual Field}.
\newblock 5 2006.

\bibitem{Freed:2006ya}
Daniel~S. Freed, Gregory~W. Moore, and Graeme Segal.
\newblock {The Uncertainty of Fluxes}.
\newblock {\em Commun. Math. Phys.}, 271:247--274, 2007.

\bibitem{Note5}
See also, e.g., \cite {Heckman:2017uxe, Bah:2020jas, Bah:2020uev, Apruzzi:2023uma, Lawrie:2023tdz} on derivation of discrete symmetry theories from 11D formulism of string theory.

\bibitem{Witten:1998wy}
Edward Witten.
\newblock {AdS / CFT correspondence and topological field theory}.
\newblock {\em JHEP}, 12:012, 1998.

\bibitem{Bergman:2022otk}
Oren Bergman and Shinji Hirano.
\newblock {The holography of duality in $ \mathcal{N} $ = 4 Super-Yang-Mills theory}.
\newblock {\em JHEP}, 11:069, 2022.

\bibitem{GarciaEtxebarria:2024jfv}
I\~naki Garc\'\i{}a~Etxebarria, Jes\'us Huertas, and Angel~M. Uranga.
\newblock {SymTFT Fans: The Symmetry Theory of 4d N=4 Super Yang-Mills on spaces with boundaries}.
\newblock 9 2024.

\bibitem{Aharony:2008ug}
Ofer Aharony, Oren Bergman, Daniel~Louis Jafferis, and Juan Maldacena.
\newblock {N=6 superconformal Chern-Simons-matter theories, M2-branes and their gravity duals}.
\newblock {\em JHEP}, 10:091, 2008.

\bibitem{Aharony:2008gk}
Ofer Aharony, Oren Bergman, and Daniel~Louis Jafferis.
\newblock {Fractional M2-branes}.
\newblock {\em JHEP}, 11:043, 2008.

\bibitem{vanBeest:2022fss}
Marieke van Beest, Dewi S.~W. Gould, Sakura Schafer-Nameki, and Yi-Nan Wang.
\newblock {Symmetry TFTs for 3d QFTs from M-theory}.
\newblock 10 2022.

\bibitem{Cordova:2018cvg}
Clay C\'ordova, Thomas~T. Dumitrescu, and Kenneth Intriligator.
\newblock {Exploring 2-Group Global Symmetries}.
\newblock {\em JHEP}, 02:184, 2019.

\bibitem{Seiberg:1996bd}
Nathan Seiberg.
\newblock {Five-dimensional SUSY field theories, nontrivial fixed points and string dynamics}.
\newblock {\em Phys. Lett.}, B388:753--760, 1996.

\bibitem{Note6}
Ethan Torres, private communication.

\bibitem{Heckman:2024oot}
Jonathan~J. Heckman, Max H\"ubner, and Chitraang Murdia.
\newblock {On the holographic dual of a topological symmetry operator}.
\newblock {\em Phys. Rev. D}, 110(4):046007, 2024.

\bibitem{Gaiotto:2009mv}
Davide Gaiotto and Alessandro Tomasiello.
\newblock {The gauge dual of Romans mass}.
\newblock {\em JHEP}, 01:015, 2010.

\bibitem{Heckman:2021vzx}
Jonathan~J. Heckman, Andrew~P. Turner, and Xingyang Yu.
\newblock {Disorder averaging and its UV discontents}.
\newblock {\em Phys. Rev. D}, 105(8):086021, 2022.

\bibitem{GarciaEtxebarria:toappear}
F.~Gagliano and I.~Garcia~Etxebarria.
\newblock Symtfts for u(1) symmetries from descent.

\bibitem{Apruzzi:2024htg}
Fabio Apruzzi, Francesco Bedogna, and Nicola Dondi.
\newblock {SymTh for non-finite symmetries}.
\newblock 2 2024.

\bibitem{Bonetti:2024cjk}
Federico Bonetti, Michele Del~Zotto, and Ruben Minasian.
\newblock {SymTFTs for Continuous non-Abelian Symmetries}.
\newblock 2 2024.

\bibitem{Cvetic:2024dzu}
Mirjam Cveti\v{c}, Ron Donagi, Jonathan~J. Heckman, Max H\"ubner, and Ethan Torres.
\newblock {Cornering Relative Symmetry Theories}.
\newblock 8 2024.

\bibitem{Bergman:2024aly}
Oren Bergman, Eduardo Garcia-Valdecasas, Francesco Mignosa, and Diego Rodriguez-Gomez.
\newblock {Non-BPS branes and continuous symmetries}.
\newblock 6 2024.

\bibitem{Heckman:2017uxe}
Jonathan~J. Heckman and Luigi Tizzano.
\newblock {6D Fractional Quantum Hall Effect}.
\newblock {\em JHEP}, 05:120, 2018.

\bibitem{Bah:2020jas}
Ibrahima Bah, Federico Bonetti, Ruben Minasian, and Peter Weck.
\newblock {Anomaly Inflow Methods for SCFT Constructions in Type IIB}.
\newblock {\em JHEP}, 02:116, 2021.

\bibitem{Bah:2020uev}
Ibrahima Bah, Federico Bonetti, and Ruben Minasian.
\newblock {Discrete and higher-form symmetries in SCFTs from wrapped M5-branes}.
\newblock {\em JHEP}, 03:196, 2021.

\bibitem{Apruzzi:2023uma}
Fabio Apruzzi, Federico Bonetti, Dewi S.~W. Gould, and Sakura Schafer-Nameki.
\newblock {Aspects of Categorical Symmetries from Branes: SymTFTs and Generalized Charges}.
\newblock 6 2023.

\bibitem{Lawrie:2023tdz}
Craig Lawrie, Xingyang Yu, and Hao~Y. Zhang.
\newblock {Intermediate defect groups, polarization pairs, and noninvertible duality defects}.
\newblock {\em Phys. Rev. D}, 109(2):026005, 2024.

\end{thebibliography}

\end{document}